\documentclass[pre,twocolumn,showpacs,aps,floatfix]{revtex4-1}

\usepackage{amsmath}
\usepackage{graphicx}
\usepackage{bm}
\usepackage{color}

\begin{document}

\title{Average-atom treatment of relaxation time in X-ray Thomson scattering from warm-dense matter.}
\author{W. R. Johnson}
 \email{johnson@nd.edu}
\affiliation{Department of Physics, 225 Nieuwland Science Hall,\\
University of Notre Dame, Notre Dame, IN 46556}
\author{J. Nilsen}
\affiliation{Lawrence Livermore National Laboratory,PO Box 808, Livermore,
CA 94551}

\date{\today}

\begin{abstract}
 The influence of finite relaxation times on Thomson scattering from warm-dense plasmas is examined within the framework of the average-atom approximation. Presently most calculations use the collision-free Lindhard dielectric function to evaluate the free-electron contribution to the Thomson cross section. In this work, we use the Mermin dielectric function, which includes relaxation time explicitly. The relaxation time is evaluated by treating the average atom as an impurity in a uniform electron gas and depends critically on the transport cross section. The calculated relaxation rates agree well with values inferred from the Ziman formula for the static conductivity and also with rates inferred from a fit to the frequency-dependent conductivity. Transport cross sections determined by the phase-shift analysis in the average-atom potential are compared with those evaluated in the commonly used Born approximation. The Born approximation converges to the exact cross sections at high energies; however, differences that occur at low energies lead to corresponding differences in relaxation rates.
The relative importance of including relaxation time when modeling X-ray Thomson scattering spectra is examined by comparing calculations of the free-electron dynamic structure function for Thomson scattering using Lindhard and Mermin dielectric functions. Applications are given to warm-dense Be  plasmas, with temperatures ranging from 2 to 32~eV and  densities ranging from 2 to 64~g/cc.
\end{abstract}


\maketitle

\section{Introduction}

 X-ray Thomson scattering \cite{GR:09} is a very promising technique for measuring the temperature, density, and ionization state in dense plasmas. Measuring these quantities is very important for understanding and modeling high energy density physics experiments. The importance of including finite relaxation times in calculations of the dynamic structure function for X-ray Thomson scattering in warm-dense plasmas is well established \cite{HR:04,HR:05,HB:07,TB:08,FW:10,PS:12}. In this paper, the influence of finite relaxation times on Thomson scattering from warm-dense plasmas is examined within the framework of the average-atom approximation. Presently most calculations use  the Lindhard \cite{Li:54} dielectric function $\epsilon^0(k,\omega)$, which describes a collision-free electron gas, to evaluate the free-electron contribution to the Thomson cross section. In this work, the Lindhard dielectric function is replaced by the \citet{NM:70} dielectric function $\epsilon^\text{M}(k,\omega)$, which includes effects of collisions and conserves the local electron number:
\begin{multline}
\epsilon^\text{M}(k,\omega) = \\ 1 + \frac{(1+i/\omega\tau)(\epsilon^0(k,\omega+i/\tau)-1)}{1 + (i/\omega\tau)(\epsilon^0(k,\omega+i/\tau)-1)/(\epsilon^0(k,0)-1)}, \label{eq1}
\end{multline}
where  $\tau$ is the relaxation time.

Schemes for including relaxation effects in the dynamic structure function have been reviewed in Ref. \cite{GR:09}. In Refs.~\cite{HR:04,HR:05,HB:07,TB:08,FW:10}, the Born approximation in a screened Coulomb potential was used to obtain $\tau$.  As an alternative, in Ref.~\cite{PS:12}, the frequency-dependent conductivity was evaluated using quantum molecular dynamics and the relaxation time was determined by fitting the frequency-dependent conductivity $\sigma(\omega)$ to the classical Drude model \cite{PD:1,PD:3},
\begin{equation}
\sigma_\text{D}(\omega) = \frac{n_I e^2}{m} \frac{\tau}{1+(\omega\tau)^2} , \label{drude}
\end{equation}
where $e$ and $m$ are the electron charge and mass and $n_I$ is the ion density.

In the present study, the relaxation time is determined using a model developed to treat scattering from impurities in a uniform electron gas \cite{ZI:72}. In this model, the relaxation time is expressed in terms of the transport cross section,  which is evaluated in the average-atom potential.

For comparison purposes,
an estimate of  $\tau$ is made by equating the average-atom version of Ziman's formula for the static conductivity \cite{JGB:06} with the static Drude conductivity $\sigma_\text{D}(0)$, which contains $\tau$ as a parameter. A somewhat different estimate of $\tau$ is obtained, following the scheme used in Ref.~\cite{PS:12}, by fitting the dynamic conductivity  $\sigma(\omega)$ to $\sigma_\text{D}(\omega)$. In the present study, $\sigma(\omega)$ is obtained from an average-atom version of the Kubo-Greenwood \cite{KU:56,KU:57,GR:58} equation.  The resulting two estimates are found to agree well with the direct calculation over a wide range of densities and temperatures.

Inasmuch as the Born approximation is widely used in calculations of transport cross sections, the present calculations are also compared with calculations carried out using the Born approximation to the transport cross section. The Born approximation converges to the exact cross section as energy increases; however, differences found at low energies lead to differences in the relaxation rates.

The utility of the average-atom approach rests on its simplicity and wide range of applicability. The present version of the average-atom model was used in Refs.~\cite{NJ:05,NJ:06,FG:06,FG:07,NC:08} to investigate anomalous dispersion in C, Al, Ag, and other plasmas in the soft X-ray region of the spectrum and has been used to investigate Thomson scattering in Refs.~\cite{SG:08,JN:12,MW:13,NJ:13,JN:14,SS:14,ST:14,SP:14,HB:15}.

In the following section, calculations of transport cross sections, relaxation times and conductivities
are described and relaxation times for Be plasmas are evaluated as functions of temperature and density.
Comparisons are made with estimates obtained from static and dynamic conductivities. Comparisons are also made with rates obtained using the Born approximation.
Finally, in Section~\ref{diel}, the Mermin dielectric function is discussed and free-electron contributions to the Thomson scattering structure function obtained using Mermin and Lindhard dielectric functions are compared for Be plasmas   with temperatures ranging from 2 to 32 eV and densities ranging from 2 to 64 g/cc.

\section{Relaxation Times and Conductivities}

The relaxation rate $\nu=1/\tau$ in the average atom picture is given by
a finite-temperature version of the impurity scattering rate,
\cite{ZI:72,BF:15}
\begin{equation}
\nu = -\frac{n_I}{m} \int_0^\infty dE \frac{df(E)}{dE}\,p\, \sigma^\text{tr}(p) , \label{nu}
\end{equation}
with $p=\sqrt{2 m E}$. In the above, $f(E)$ is the Fermi distribution function
\begin{equation}
 f(E) = \frac{1}{1+ \exp[\beta(E-\mu)]},
\end{equation}
where $\beta = 1/k_\text{\tiny B} T$ and $\mu$ is the chemical potential.
In Eq.~(\ref{nu}), $\sigma^\text{tr}(p)$ is the transport cross section,
which is given in terms of scattering phase shifts $\delta_l(p)$ by
\begin{equation}
\sigma^\text{tr}(p)  = \frac{4\pi}{p^2} \sum_{l=0}^\infty (l+1) \sin^2(\delta_{l+1}(p)-\delta_l(p)) . \label{ph}
\end{equation}

\begin{figure}
\centerline{\includegraphics[scale=0.68]{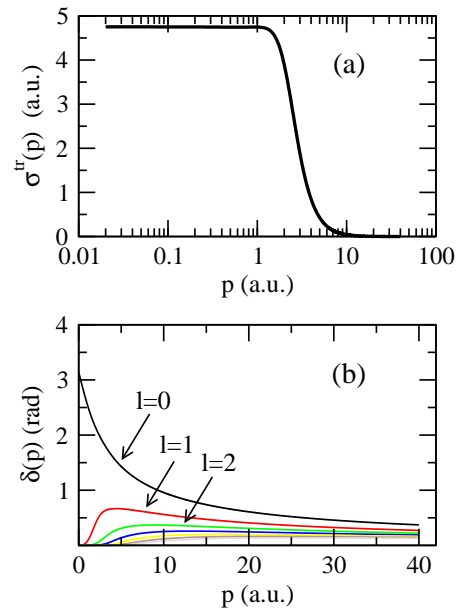}}
\caption{\label{fig1} Color online. (a) Transport cross section $\sigma^\text{tr}(p)$ for a warm-dense  Be plasma at density $\rho=8$~g/cc and  $T=16$~eV.  (b) Electron-ion scattering phase shifts $\delta(p)$ in the average-atom potential used to calculate the cross section shown in panel (a).}
\end{figure}

In panel (a) of Fig.~\ref{fig1}, the transport cross section $\sigma^\text{tr}(p)$ is illustrated for a Be plasma at temperature $T=16$~eV and density $\rho=8$~g/cc.
The ion potential is taken to be the finite-range average-atom potential $V(r)$:
$V(r)\to -Ze^2/r$ as $r\to 0$ and $ V(r) = 0$ for $r \geq R_\text{\tiny WS}$, where
$R_\text{\tiny WS}$ is the Wigner-Seitz radius.  In this example $Z=4$ and
$R_\text{\tiny WS} = 1.444\ a_0$. In panel (b) of Fig.~\ref{fig1}, we show the phase-shifts $\delta_l(p)$ for continuum states in the average-atom potential $V(r)$. These phase shifts are used in Eq.~(\ref{ph})  to obtain the cross section shown in panel (a) of Fig.~\ref{fig1}.
Since the K-shell of the Be ion is occupied in the average-atom model under the present conditions of temperature and density, the fact that the s-wave phase-shift $\delta_0(0)=\pi$, whereas $\delta_l(0)=0$ for $l>0$, is in harmony with Levinson's theorem.
The ion density in the present example is $n_I=0.07932$~$a_0^{-3}$ and the chemical potential is $\mu=1.160$~a.u..
The relaxation frequency from Eq.~(\ref{nu}) is $\nu =0.4510$~a.u. and the corresponding relaxation time is $\tau=2.217\ \text{a.u.}$.
(Note that a.u.\ refers to atomic units in which $e=\hbar=m=1$, with 1 a.u.\ in time $= 2.4189\times 10^{-17}$~s and 1 a.u.\ in cross section $= 2.800\times 10^{-17}$~cm$^2$.)

In Table~\ref{tab1}, we list values of effective ionic charge $Z^\ast$, chemical potential $\mu$, relaxation rate $\nu$ and relaxation time $\tau$ for a Be plasma at temperatures ranging from $T=2$~eV to 32~eV and densities ranging from $\rho=2$~g/cc to 64~g/cc. For each fixed temperature, the relaxation time decreases systematically with density.

In the following subsection, we compare
the rates obtained from Eq.~(\ref{nu}) with rates estimated from conductivity calculations.
\begin{table}
 \caption{Properties of warm-dense Be at density $\rho$ and temperature $T$:  $Z^\ast$ effective ionic charge,
 $\mu$ chemical potential, $\nu$ collision rate, $\tau$ relaxation time are given in a.u.. \label{tab1}}
 \begin{center}
\begin{tabular}{ccccc}
\hline\hline
$\rho$~(g/cc)  & $Z^\ast$ &    $\mu$  &   $\nu$  & $\tau$ \\
\hline
\multicolumn{5}{c}{$T=2$ eV} \\							
2	&	1.406	&	0.429	&	0.149	&	6.713	\\
4	&	1.813	&	0.822	&	0.313	&	3.196	\\
8	&	2.124	&	1.457	&	0.585	&	1.710	\\
16	&	2.379	&	2.498	&	0.974	&	1.026	\\
32	&	2.610	&	4.219	&	1.411	&	0.709	\\
64	&	2.833	&	7.074	&	1.930	&	0.518	\\
\multicolumn{5}{c}{$T=4$ eV} \\
2	&	1.406	&	0.393	&	0.136	&	7.351	\\
4	&	1.798	&	0.800	&	0.298	&	3.350	\\
8	&	2.115	&	1.443	&	0.578	&	1.729	\\
16	&	2.375	&	2.490	&	0.975	&	1.026	\\
32	&	2.608	&	4.214	&	1.428	&	0.700	\\
64	&	2.832	&	7.071	&	1.818	&	0.550	\\
\multicolumn{5}{c}{$T=8$ eV} \\
2	&	1.455	&	0.250	&	0.109	&	9.162	\\
4	&	1.763	&	0.708	&	0.256	&	3.910	\\
8	&	2.084	&	1.388	&	0.550	&	1.818	\\
16	&	2.361	&	2.457	&	0.970	&	1.031	\\
32	&	2.603	&	4.196	&	1.436	&	0.696	\\
64	&	2.830	&	7.060	&	1.813	&	0.552	\\
\multicolumn{5}{c}{$T=16$ eV} \\
2	&	1.581	&	-0.241	&	0.071	&	14.09	\\
4	&	1.746	&	0.359	&	0.178	&	5.609	\\
8	&	2.009	&	1.160	&	0.451	&	2.217	\\
16	&	2.307	&	2.320	&	0.928	&	1.077	\\
32	&	2.581	&	4.118	&	1.460	&	0.685	\\
64	&	2.823	&	7.018	&	1.842	&	0.543	\\
\multicolumn{5}{c}{$T=32$ eV} \\
2	&	1.903	&	-1.617	&	0.044	&	22.83	\\
4	&	1.918	&	-0.694	&	0.106	&	9.401	\\
8	&	2.030	&	0.401	&	0.288	&	3.475	\\
16	&	2.247	&	1.817	&	0.733	&	1.365	\\
32	&	2.526	&	3.815	&	1.436	&	0.696	\\
64	&	2.800	&	6.849	&	1.956	&	0.511	\\
\hline\hline								 											 			 \end{tabular}
 \end{center}
\end{table}

\subsection{Estimates from Conductivity Calculations}
An average atom version of the Kubo-Greenwood (KG) equation for the frequency-dependent conductivity was derived in \cite{JGB:06} by considering linear response of the average atom to a time-varying electric field:
\begin{equation}
\sigma(\omega) = \frac{2 n_I \pi e^2}{m \omega} \sum_{ij} (f_i-f_j)\,
|\langle j| p_z |i \rangle |^2\,
 \delta(\epsilon_j-\epsilon_i-\omega).
\end{equation}
In this equation, $n_I$ is the ion density, $\epsilon_i$ and $f_i$ are the energy and Fermi distribution function of average-atom state $i$, and $p_z$ is the $z$ component of the momentum operator.
Contributions to the conductivity arise from three distinct processes:  free-free transitions, bound-bound transitions (discrete spectra) and bound-free transitions (photoionization).  The free-free contribution to the conductivity, which diverges at low frequencies,  was regulated in an ad-hoc manner in Ref.~\cite{JGB:06}. It was later shown \cite{KJ:08} how the free-free contribution to the KG equation could be reformulated to include finite collision times and that the resulting free-free contribution was regular at $\omega=0$. The modified free-free contribution to the KG equation is given by a frequency-dependent generalization of the Ziman formula
\begin{equation}
\sigma(\omega) = - \frac{2e^2}{3 m^2} \int \frac{d^3p}{(2\pi)^3}
 \left(\frac{\partial f}{\partial E}\right) p^2 \frac{\tau_p}{1+\omega^2\tau_p^2} . \label{zi}
\end{equation}
It should be emphasized that $\tau_p$ in Eq.~(\ref{zi}) is the mean time between collisions for an electron with momentum $p$ [$\tau_p = V/(p \sigma^{tr}(p))$],  not the frequency independent parameter $\tau = 1/\nu$. Indeed,
 $\tau_p$ can be determined from the mean-free-path $\Lambda_p$ by $\tau_p = \Lambda_p/v$, where $v$ is
the electron velocity.  The mean-free-path is related to
the transport cross-section $\sigma^\text{tr}(p)$ by $\Lambda_p=V/\sigma^\text{tr}(p)$, where $V = 1/n_I$ is the volume of the WS cell. In the static limit, $\sigma(\omega)$ reduces to
the Ziman formula \cite[Eq.~(7.25)]{ZI:72}.
Discussions of conductivity in the average-atom approximation can be found in Refs.~\cite{JGB:06,KJ:08,WJ:09}, while comparisons of average-atom conductivities with experiment and with other calculations are found in \cite{WJ:09,EJ:11,SC:12}.

In column~3 of Table~\ref{tab2}, we compare estimates of relaxation rates $\nu_1$, obtained
by equating $\sigma(0)$ from Eq.~(\ref{zi}) to $\sigma_D(0)$ from Eq.~(\ref{drude}), with
values of $\nu$ obtained from Eq.~(\ref{nu}).

\begin{figure}
\centerline{\includegraphics[scale=0.68]{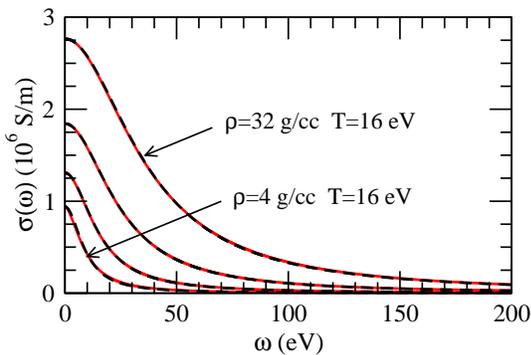}}
\caption{Color online. Frequency-dependent conductivities $\sigma(\omega)$ calculated using the Ziman formula Eq.~(\ref{zi})  are shown in the solid
red lines for warm-dense Be plasmas with densities $\rho = (4,8,16,32)$~g/cc and temperature $T=16$~eV.
The lower curves correspond to lower densities. The
dashed black lines represent one parameter ($\nu_2$) fits of the conductivities to the Drude formula (\ref{drude}). Values of the collision rates $\nu_2$ obtained from the fits are compared with those obtained from Eq.(\ref{eq8}) in Table~\ref{tab2}. \label{fig2}}
\end{figure}

In Fig.~\ref{fig2}, we show the free-free contributions to frequency-dependent conductivities $\sigma(\omega)$ from Eq.~(\ref{zi}) for Be plasmas at temperature $T=16$~eV  and densities $\rho$ ranging from 4 to 32~g/cc
in the solid red lines. The dashed black lines show results
of one-parameter ($\nu_2$) fits of $\sigma(\omega)$ to the
function
\begin{equation}
  \frac{\sigma(0)}{1+(\omega/\nu_2)^2} ,  \label{eq8}
\end{equation}
which is obtained from Eq.~(\ref{drude}) by requiring $\sigma_\text{D}(0) = \sigma(0)$.

As can be seen from Fig.~\ref{fig2}, the resulting fit reproduces $\sigma(\omega)$ accurately for the cases considered.  Estimates of relaxation rates $\nu_2$ for Be over a range of temperatures and densities are listed in column 4 of Table~\ref{tab2}. As can be seen from the table, there is good agreement between
the estimates $\nu_1$ and $\nu_2$ and agreement to better than a factor of 2 between the direct calculation of $\nu$ and the two estimates.

\begin{table}
\caption{Relaxation rates $\nu$ for Be plasmas at temperature $T$ and
density $\rho$ are compared with rates
$\nu_1$ inferred from static conductivity calculations,
with rates $\nu_2$ obtained by fitting dynamic conductivity calculations and with rates $\nu_\text{B}$ obtained by using
the Born approximation to evaluate the transport cross section. Units for relaxation rates are a.u..
\label{tab2}}
\begin{center}
\begin{tabular}{ccccc}
\hline\hline
\multicolumn{1}{c}{$\rho$~g/cc}    &
\multicolumn{1}{c}{$\nu$}          &
\multicolumn{1}{c}{$\nu_1$}        &
\multicolumn{1}{c}{$\nu_2$}        &
\multicolumn{1}{c}{$\nu_\text{B}$} \\
 \hline
\multicolumn{5}{c}{$T=4$~eV}                        \\
4	&	0.298	&	0.318	&	0.314	&	0.839	\\
8	&	0.578	&	0.585	&	0.584	&	0.952	\\
16	&	0.975	&	0.971	&	0.971	&	1.067	\\
32	&	1.428	&	1.423	&	1.422	&	1.174	\\
\multicolumn{5}{c}{$T=8$~eV}	 	 	 		 	\\
4	&	0.256	&	0.327	&	0.320	&	0.741	\\
8	&	0.550	&	0.579	&	0.578	&	0.934	\\
16	&	0.970	&	0.958	&	0.958	&	1.069	\\
32	&	1.436	&	1.409	&	1.407	&	1.176	\\
\multicolumn{5}{c}{$T=16$~eV} 	 	 	 		    \\
4	&	0.178	&	0.338	&	0.333	&	0.486	\\
8	&	0.451	&	0.561	&	0.559	&	0.793	\\
16	&	0.928	&	0.913	&	0.911	&	1.039	\\
32	&	1.460	&	1.361	&	1.354	&	1.181	\\
 \multicolumn{5}{c}{$T=32$~eV}	 	 	 		    \\
4	&	0.106	&	0.297	&	0.289	&	0.253	\\
8	&	0.288	&	0.488	&	0.480	&	0.504	\\
16	&	0.733	&	0.790	&	0.780	&	0.844	\\
32	&	1.436	&	1.212	&	1.192	&	1.129	\\
\hline\hline
\end{tabular}
\end{center}
\end{table}

\subsection{Born approximation}
\begin{figure}
\centerline{\includegraphics[scale=0.68]{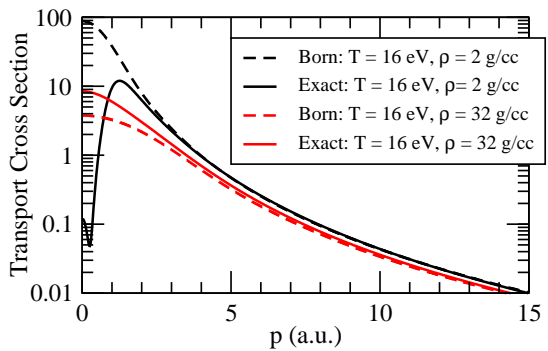}}
\caption{\label{fig3} Color online. Exact transport cross sections for a Be plasma at $T=16~eV$ (shown in solid lines) are compared with the Born approximation (shown in dashed lines). At density $\rho=2$~g/cc (shown in black lines), the Born approximation cross section is much larger than the exact cross section (shown in red lines) at small values of momentum $p$, while at $\rho=32$~g/cc, the Born approximation is smaller at small $p$. In both cases, the Born approximation approaches the exact cross section as $p$ increases.}
\end{figure}

In many studies of relaxation rates \cite{HR:04,HB:07,TB:08,FW:10},  the scattering cross section $\sigma_p(\theta)$ is evaluated using the Born approximation for the
scattering amplitude $f(\theta)$:
\begin{equation}
     f(\theta) = -\frac{1}{2\pi} V(q) , \label{born}
\end{equation}
where $V(q)$ is the Fourier transform of the electron-ion scattering potential V(r),
which is typically assumed to be an exponentially damped ion potential.
In the above, $\bm{q} = \bm{p}_1 -\bm{p}_2$ is the momentum transferred to the ion.
We have omitted a factor of $m/\hbar$ in Eq.~(\ref{born}) since we use atomic units.
For elastic scattering $p_2=p_1=p$ and $q^2 = 2 p^2(1-\mu)$, where $\mu$ is the cosine of the angle between $\bm{p}_1$ and $\bm{p}_2$. The Fourier transform of the potential $V(r)$ is given by
\begin{equation}
V(q) =  \frac{4\pi}{q} \int_0^\infty\!\! r \sin(q r) V(r)\, dr,
\end{equation}
and the transport cross section is
\begin{equation}
\sigma^\text{tr}(p) = \frac{1}{4\pi p^4}\int_0^{2p}\!\! q^3\, |V(q)|^2\, dq .
\end{equation}
In Fig.~\ref{fig3}, we compare the Born-approximation transport cross sections for a Be plasma at $T=16$~eV, which are shown in dashed lines,  with ``exact" cross sections, which are shown in solid lines. The average-atom
potential is used to evaluate the the Born cross section.
 For small values of $p$, the Born cross section for density $\rho=2$~g/cc is larger than the exact cross section, while for $\rho=32$~g/cc, the Born cross section is smaller than the exact cross section. As $p$ increases, the Born approximation approaches the exact cross section in both cases.
Differences between the Born and exact cross sections at small values of $p$ are reflected in the relaxation rates. Thus,  for $T=16$~eV and $\rho= (2,\ 32)$~g/cc, Eq.~(\ref{nu}) gives  $\nu =  (0.0710,\ 1.460)$~a.u.\ , using the exact transport cross section, whereas Eq.~(\ref{nu}) gives $\nu_\text{B} = (0.2433,\  1.181)$~a.u.\ using the Born transport cross section. Born approximation calculations of relaxation rates in Be plasmas for other values of temperature and  density are compared with direct calculations and with values inferred from conductivity calculations in Table~\ref{tab2}.

\subsection{Comparisons}
\citet{PS:12} obtain frequency independent values of $\nu$
by fitting the frequency-dependent conductivity obtained from a quantum molecular dynamics
calculation to the Drude model. For uncompressed Be at density 1.85~g/cc and temperature 12~eV and compressed Be at density 5.5~g/cc and temperature 13~eV,
values $\nu = 0.357$ and 0.703~a.u., respectively, were obtained. The present calculation using Eq.~\ref{nu}  predicts  values $\nu = 0.0800$ and 0.317~a.u., respectively. The value of $\nu$ at $\rho=1.85$~g/cc is substantially smaller than that value predicted in Ref.~\cite{PS:12}.

The ``mathematical" reason for the relatively small value of $\nu$ at metallic density
is strong interference at small values of $p$ between partial waves with $l=0$ and $l=1$ in the expression for $\sigma^\text{tr}(p)$, resulting in a corresponding reduction in the
value of the relaxation frequency. This reduction is evident in the black curve in Fig.~\ref{fig3}, where the partial-wave expression for the transport cross section at $\rho=2$~g/cc and $T=16$~eV is compared with the Born approximation. One consequence of the interference, pointed out earlier, is a substantial increase in the size of $\nu_\text{B}$ relative to $\nu$; a second consequence is that substantial differences arise between relaxation frequencies $\nu,\ \nu_1$ and $\nu_2$.
In that regard,
average-atom relaxation rates obtained  by fitting frequency-dependent conductivities to the Drude formula $\nu_2 = 0.244$ and 0.426 are somewhat closer to the values obtained in Ref.~\cite{PS:12}.  Differences between the respective values of $\nu_2$ reflect differences between average-atom and QMD calculations of frequency-dependent conductivities.

\citet{FB:16} determined  values of $\nu$ for Be at metallic density and temperature $T=10$~eV within the framework of the SCAALP \cite{FB:12} average-atom model . The value $\nu_\text{KG}=0.153$~a.u. was obtained using the Kubo-Greenwood theory;  the corresponding value from the present calculation is $\nu_2=0.246$.
Furthermore, $\nu_\text{B}= 0.363$~a.u.\ was obtained in \cite{FB:16} using an expression for the frequency-dependent conductivity based on the Born approximation, and $\nu_{LB}=0.278$~a.u. was obtained using the Lenard-Balescu theory together with the average-atom electron-ion potential.
The present calculation gives $\nu = 0.0896$~a.u.\ for $\rho=1.85$~g/cc and $T=10$~eV which is again much smaller than the values obtained in \cite{FB:16}
for reasons mentioned in the previous paragraph.
It should be mentioned that the average-atom Born approximation result gives $\nu_\text{B}=0.369$, in good agreement with the value $\nu_\text{B}=0.363$ obtained in \cite{FB:16}.

\section{Dielectric Function\label{diel}}

\begin{figure}
\centerline{\includegraphics[scale=0.68]{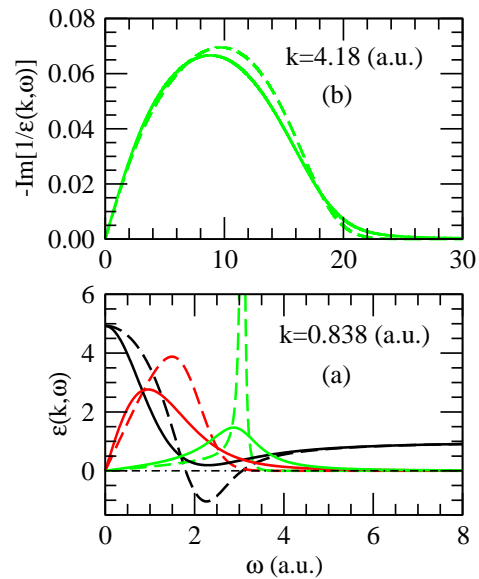}}
\caption{\label{fig4} Color online. Contributions to dielectric functions of a Be plasma at
temperature $T=16$~eV and density $\rho = 8$~g/cc. The collision rate
 is $\nu=0.4510$~a.u..  Dashed and solid lines illustrate contributions to the Lindhard and Mermin dielectric functions, respectively.  (a)
$\text{Re}[\epsilon(k,\omega)]$ (black lines), $\text{Im}[\epsilon(k,\omega)]$
(red lines) and -$\text{Im}[1/\epsilon(k,\omega)]$ (green lines) are shown for $k=0.838$~a.u. corresponding to scattering of a 9000~eV X-ray at angle $20^\circ$.
   (b) -$\text{Im}[1/\epsilon(k,\omega)]$ for $k=4.18$~a.u. corresponding to scattering of a 9000~eV X-ray at $120^\circ$.}
\end{figure}

Expressions for the Lindhard and Mermin dielectric functions are reduced to  single integrals suitable for numerical studies in Appendix~\ref{appa}.
In Fig.~\ref{fig4}, we compare these two functions of $\omega$ at fixed values of $k$ for a Be plasma at temperature $T=16$~eV and density $\rho = 8$~g/cc. The corresponding collision rate, given in Table~\ref{tab1} is $\nu=0.4510$~a.u..
The lower panel of the figure shows the real and imaginary parts for the Lindhard and Mermin dielectric functions together with the function -$\text{Im}[1/\epsilon(k,\omega)]$, which occurs in the expression for free-electron dynamic structure function.
These functions are evaluated at $k=0.838$~a.u., corresponding to scattering of a 9000~eV X-ray at $20^\circ$.
The Lindhard function -$\text{Im}[1/\epsilon(k,\omega)]$, shown in the dashed green line, resonates (plasmon resonance) near the second zero of $\text{Re}[\epsilon(k,\omega)]$, shown by the dashed black line. This resonance is seen to be strongly damped in the Mermin dielectric function shown in the solid green line.
The upper panel of Fig~\ref{fig2} compares Lindhard and Mermin calculations of $-\text{Im}[1/\epsilon(k,\omega)]$ at $k=4.18$~a.u. corresponding to scattering a 9000~eV Xray at $120^\circ$. The resulting values of $\text{Re}[\epsilon(k,\omega)]$ are close to 1 and those of  $\text{Im}[\epsilon(k,\omega)]$ are indistinguishable from the green lines
in the figure. It is particularly interesting to note that the maximum  of $\text{Im}[\epsilon(k,\omega)]$ is reduced in amplitude and shifted to lower energies in the Mermin calculation.

\section{Thomson Scattering\label{thom}}

We now turn to applications of the Mermin dielectric function to Thomson scattering. The contribution to the dynamic structure function $S(k,\omega)$ from inelastic scattering by free electrons is given by
\begin{equation}
S_{ee}(k,\omega) = -\frac{1}{1-\exp(-\omega/T)}\frac{Z^\ast k^2}{4\pi n_e} \text{Im}\left[\frac{1}{\epsilon(k,\omega)} \right].
\end{equation}
It should be noted that $Z^\ast/n_e = V$, the volume of an average-atom Wigner-Seitz cell.

\begin{figure}
\centerline{\includegraphics[scale=0.4]{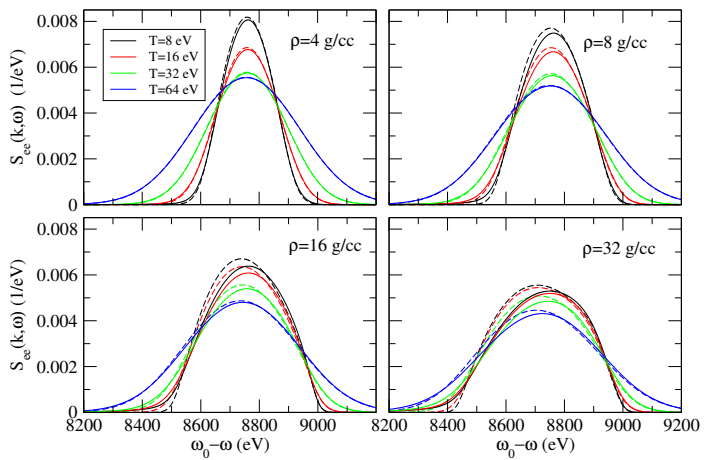}}
\caption{\label{fig5} Color online. Free-electron dynamic structure functions $S_{ee}(k,\omega)$ for Thompson scattering of 9000~eV X-rays at $120^\circ$ from Be plasmas with temperatures ranging from 8 to 64~eV and plasma densities ranging from 4 to 32~g/cc. The dashed curves describe calculations done using the Lindhard dielectric function and the solid curves represent those done using the Mermin dielectric function. The curves with smaller amplitudes describe plasmas with higher density at a given temperature
and with higher temperature at a given density. }
\end{figure}

In Fig.~\ref{fig5}, we compare free-electron dynamic structure functions for Thompson scattering of 9000~eV X-rays at $120^\circ$ from Be plasmas with temperatures ranging from 8 to 64 eV and densities ranging from 4 to 32 g/cc.  The solid curves describe calculations done including relaxation time and the dashed curves represent those done ignoring relaxation effects. The maxima of the curves evaluated using the Mermin dielectric function are shifted to higher energy and reduced in amplitude compared to those evaluated using the Lindhard dielectric function.

The size of the shift increases with density at a fixed temperature and decreases with temperature at a fixed density. Since the peak of structure function is downshifted from the incident photon energy $\omega_0$ by approximately $2(\omega_0/c)^2 \sin^2(\theta/2)$~a.u., the effect of including relaxation time is similar to ignoring relaxation time and decreasing the scattering angle. (In this regard it should be noted that the horizontal axis in Fig.~\ref{fig4} is $\omega$ and in Fig.~\ref{fig5} is $\omega_0-\omega$.)

\begin{figure}
\centerline{\includegraphics[scale=0.68]{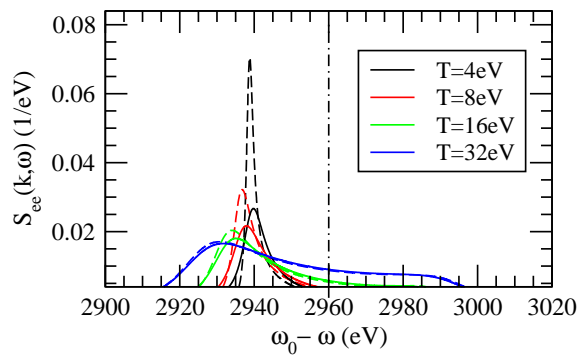}}
\caption{\label{fig6} Color online. Free-electron dynamic structure functions $S_{ee}(k,\omega)$ for Thompson scattering of 2960~eV X-rays at $40^\circ$ from Be plasmas with temperatures ranging from 4 to 32~eV and metallic density. The dashed curves describe calculations done using the Lindhard dielectric function and the solid curves represent those done using the Mermin dielectric function.}
\end{figure}

In Fig.~\ref{fig6}, we compare $S_{ee}(k,\omega)$ for scattering of 2960~eV X-rays at $40^\circ$ from Be plasmas with temperatures ranging from 4 to 32 eV and density 1.84~g/cc.  The solid curves describe calculations done including relaxation time and the dashed curves represent those done ignoring relaxation effects. As in the previous examples, maxima of the curves evaluated including relaxation effects are shifted to higher energy and reduced in amplitude compared to those in which relaxation is ignored. The size of the shift increases with density at a fixed temperature and decreases with temperature at a fixed density. The relative importance of including finite relaxation times in the coherent scattering regime (at small momentum transfer) becomes obvious on comparing Figs.~\ref{fig5} and \ref{fig6}.

\section{Summary and Conclusions}

The influence of finite relaxation times on the free electron contribution to X-ray Thomson scattering
is examined within the framework of the average-atom theory.
For this purpose, the Lindhard dielectric function, which describes a collision-free electron gas, was replaced by the Mermin dielectric function, which includes the relaxation time and conserves the local electron number. The relaxation time used in the Mermin function was obtained by treating the average atom as an impurity in a uniform electron gas.
The relaxation rate depends crucially on the transport cross section, which is evaluated in terms of phase shifts in the average-atom potential.

Examples are given for Be plasmas with temperatures ranging from 2 to 32~eV
and densities from 2 to 64~g/cc.
Rates determined from the impurity scattering formula are found to agree within a factor of 2 with rates inferred from conductivity calculations. Average-atom calculations of the conductivity $\sigma(\omega)$  are carried out using a frequency-dependent generalization of the Ziman formula. One scheme for determining relaxation times is to equate $\sigma(0)$ to the static Drude conductivity $\sigma_D(0)$, which is proportional to $\tau$.
A second scheme is to fit the frequency dependence of $\sigma(\omega)$  to the frequency-dependent Drude model.
Results from these two methods agree well with one another.

The transport cross section used in the calculation of the relaxation rate was compared with the Born approximation
cross section, often used in calculations of relaxation rates.
Significant differences between the Born and exact cross sections were
found at low momenta, leading to corresponding differences in relaxation rates.

The Mermin function $-\text{Im}[1/\epsilon^M(\omega,k)]$, which governs the free-electron contribution to Thomson scattering, was compared with its counterpart using the Lindhard dielectric function $\epsilon^0(\omega,k)$. Plasmon resonance features that show up in calculations based on the Lindhard dielectric function
at small values of $k$ were significantly broadened using the Mermin dielectric; moreover, the Compton feature that shows up at large values of $k$ was reduced in amplitude and shifted to lower energy.

Finally, plots of the free-electron contribution to the Thomson scattering structure function are presented for warm-dense Be plasmas over a range of temperatures and densities. These plots illustrate that effects of finite relaxation times are most important for low temperatures at fixed density
and for high density at fixed temperature.

\begin{acknowledgments}
The authors are grateful to K. T. Cheng, T. D\"{o}ppner and D. Krause for helpful discussions. The work of one author (J.N.) was performed under the auspices of the U.S. Department of Energy by Lawrence Livermore National Laboratory under Contract DE-AC52-07NA27344.
\end{acknowledgments}

\appendix
\section{Reduction of the Mermin Dielectric Function\label{appa}}
The Mermin dielectric function is expressed in terms of the Lindhard dielectric function
\begin{multline}
\epsilon^0(k,\omega+i\nu)\\
 = 1 -\frac{1}{\pi^2 k^2} \int \frac{f(\bm{p} + \bm{k}/2) - f(\bm{p} - \bm{k}/2)}{\bm{k}\cdot\bm{p} -\omega - i \nu} d^3p ,
\end{multline}
with frequency $\omega$ replaced by $\omega+i\nu$, where $\nu=1/\tau$ is the collision rate.
The real and imaginary parts of the Lindhard function can be reduced to the following
integrals
\begin{multline}
\text{Re}\left[\epsilon^0(k,\omega+i\nu)\right]  =  1 \\
+ \frac{2}{\pi k^3} \int_0^\infty\!\! p f(p)\, dp \left\{  \log \left| \frac{k^2 + 2 \omega  + 2 p k + 2i\nu }{k^2 + 2 \omega  -2 p k + 2i\nu }\right| \right. \\
\left. + \log \left| \frac{k^2  - 2 \omega + 2 p k - 2 i \nu}{k^2  - 2 \omega - 2 p k - 2 i \nu}\right| \right\} ,
\end{multline}
and
\begin{multline}
\text{Im}\left[\epsilon^0(k,\omega+i\nu)\right]  =  \\
\frac{2}{\pi k^3} \int_0^\infty\!\! p f(p)\, dp \left\{
 \arctan(2\nu, k^2 + 2 \omega  + 2 p k) \right. \\ -  \arctan(2\nu, k^2 + 2 \omega  - 2 p k) \\
 -  \arctan(2\nu,k^2-2\omega+2pk)\\
\left. + \arctan(2\nu, k^2  - 2 \omega - 2 p k) \right\}. \label{arc}
\end{multline}
In Eq.~(\ref{arc}), $\arctan(y,x)$ is the phase of the complex number $(x+iy)$.
In the limit $\nu\to 0$, the sum of the terms in braces in Eq.~(\ref{arc}) is $\pi$ for $|k^2-2\omega|/2k \leq p \leq (k^2+ 2\omega)/2k$ and 0 otherwise.
The Lindhard dielectric function is the limiting value as $\nu \to 0$ of  $\epsilon^0(k,\omega+i\nu)$:
\begin{multline}
\text{Re}[\epsilon^0(k,\omega)] = 1 + \\ \frac{2}{\pi k^3} \int_0^\infty\!\! p f(p)\, dp \left\{  \log \left| \frac{k^2 + 2 \omega  + 2 p k}{k^2 + 2 \omega  -2 p k}\right|  \right. \\
  \left. + \log \left| \frac{k^2  - 2 \omega + 2 p k}{k^2  - 2 \omega - 2 p k}\right| \right\} ,
 \end{multline}
\begin{multline}
\text{Im}[\epsilon^0(k,\omega)] = \frac{2}{k^3} \int_a^b p f(p) dp \\
= \frac{2 k_B T}{k^3} \log \left[\frac{1+\exp[(\mu-a^2/2)/k_BT]}{1+\exp[(\mu-b^2/2)/k_BT]}\right],
\end{multline}
where $a = |k^2-2\omega|/2k$ and $b =(k^2+ 2\omega)/2k$.


%

\end{document}